\documentclass[crystals,article,submit,moreauthors,pdftex]{mdpi} 

\firstpage{1} 
\pubvolume{1}
\issuenum{1}
\articlenumber{0}
\pubyear{2021}
\copyrightyear{2020}
\datereceived{} 
\dateaccepted{} 
\datepublished{} 
\hreflink{https://doi.org/} % If needed use \linebreak

\pdfoutput=1

\newcommand{\gc}{$^\circ$C}
\newcolumntype{Y}{>{\centering\arraybackslash}X}

\Title{Crystal Growth of the Quasi-2D Quarternary Compound AgCrP\textsubscript{2}S\textsubscript{6} by Chemical Vapor Transport}

\TitleCitation{Crystal Growth of the Quasi-2D Quarternary Compound AgCrP\textsubscript{2}S\textsubscript{6} by Chemical Vapor Transport}

\Author{Sebastian Selter $^{1}$\orcidA{}, Yuliia Shemerliuk $^{1}$, Bernd B\"{u}chner $^{1,2}$\orcidC{} and Saicharan Aswartham $^{1,}$*\orcidD{}}

\AuthorNames{Sebastian Selter, Yuliia Shemerliuk, Bernd B\"{u}chner and Saicharan Aswartham}

\AuthorCitation{Selter, S.; Shemerliuk, Y.; B\"{u}chner, B.; Aswartham, S.}

\address{%
$^{1}$ \quad Institute for Solid State Research, Leibniz IFW Dresden, Helmholtzstr. 20, 01069 Dresden, Germany\\
$^{2}$ \quad Institute of Solid State and Materials Physics and W\"urzburg-Dresden Cluster of Excellence
ct.qmat, Technische Universit\"at Dresden, 01062 Dresden, Germany}

\corres{Correspondence: s.aswartham@ifw-dresden.de}

\abstract{We report optimized crystal growth conditions for the quarternary compound AgCrP$_2$S$_6$ by chemical vapor transport. Compositional and structural characterization of the obtained crystals were carried out by means of energy-dispersive X-ray spectroscopy and powder X-ray diffraction. AgCrP$_2$S$_6$ is structurally closely related to the $M_2$P$_2$S$_6$ family, which contains several compounds that are under investigation as 2D magnets. As-grown crystals exhibit a plate-like, layered morphology as well as a hexagonal habitus. AgCrP$_2$S$_6$ crystallizes in monoclinic symmetry in the space group $P2/a$ (No.~13). The successful growth of large high-quality single crystals paves the way for further investigations of low dimensional magnetism and its anisotropies in the future and may further allow for the manufacturing of few-layer (or even monolayer) samples by exfoliation.}


\begin{document}

\section{Introduction}

Among the magnetic quasi-two-dimensional materials that have recently moved in the focus of (quasi-)two-dimensional (2D) materials research~\cite{NatNano2018,Gibertini2019,Samarth2017}, the $M_2$P$_2$S$_6$ class of layered materials offers a plenitude of isostructural compounds with different magnetic properties depending on $M$~\cite{RBrec1986,Susner2017}. Thus, $M_2$P$_2$S$_6$ allows to investigate fundamental aspects of low dimensional magnetism and several members may be promising for future applications, \textit{e.g.}, complementing non-magnetic (quasi-)2D materials in heterostructures or in spintronic devices~\cite{Zhong2017,Song2019}. 

Regarding the crystal structure, the $M_2$P$_2$S$_6$ family consists of van der Waals layered compounds which share a honeycomb network of $M^{2+}$ and, most prominently, a dominantly covalent [P$_2$S$_6$]$^{4-}$ anion located in the voids of the honeycomb~\cite{RBrec1986,Susner2017}. In the bulk, such layers are stacked on top of each other only interacting \textit{via} weak van der Waals forces. Consequently, these compounds can be easily exfoliated potentially down to a single layer~\cite{Lee2016,Kim2019}.

\begin{figure}[htb]
\centering
\includegraphics[width=\columnwidth]{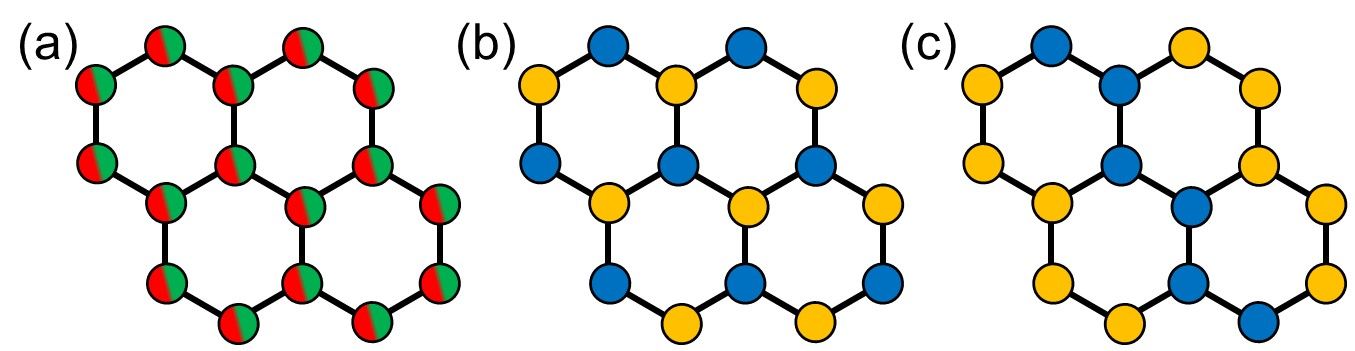}
\caption{
Schematic illustration of the different arrangements of $M$ and $M'$ on the honeycomb lattice of $M_2$P$_2$S$_6$. (a) Random distribution for $M^{2+}M'^{2+}$P$_2$S$_6$. (b) Alternating/triangular arrangement and (c) zig-zag stripe like arrangement for $M^{1+}M'^{3+}$P$_2$S$_6$.
}
\label{fig:Magn_Struc_Scheme}
\end{figure}

\begin{sloppypar}Several isovalent substitution series of $M^{2+}$ by another $M'^{2+}$ (\textit{e.g.}, (Mn$_{1-x}$Fe$_x$)$_2$P$_2$S$_6$~\cite{TMasubuchi2008}, (Mn$_{1-x}$Ni$_x$)$_2$P$_2$S$_6$~\cite{Shemerliuk2021}, (Fe$_{1-x}$Ni$_x$)$_2$P$_2$S$_6$~\cite{RRao1992,Selter2021} and (Zn$_{1-x}$Ni$_x$)$_2$P$_2$S$_6$~\cite{RBrec1986}) are reported to exhibit solid solution behavior and, thus, imply a random distribution of the substituents on the honeycomb network, as illustrated in Fig.~\ref{fig:Magn_Struc_Scheme}(a). Beyond isovalent substitution, Colombet~\textit{et al.}~\cite{PColombet1982,Ouili1987,PColombet1983,Lee1986} demonstrated that a substitution of $M_2^{2+}$ by $M^{1+}M'^{3+}$ also yields several stable compounds. In contrast to the isovalent substitution series however, $M^{1+}$ and $M'^{3+}$ do not randomly occupy the $M$ positions in the lattice but order either in an alternating or in a zig-zag stripe-like arrangement on the honeycomb, as illustrated in Fig.~\ref{fig:Magn_Struc_Scheme}(b) and (c), respectively. The former arrangement is attributed to a minimization of repulsive Coloumb interactions (\textit{i.e.}, charge ordering). The latter is observed for compounds for which $M^{1+}$ and $M'^{3+}$ have notably different sizes and, thus, is dominantly driven by a minimization of lattice distortion and steric effects~\cite{PColombet1982,RBrec1986}.\end{sloppypar}

With $M'^{3+}$ being a magnetic ion (\textit{e.g.}, V$^{3+}$ or Cr$^{3+}$) and $M^{1+}$ being non-magnetic (\textit{e.g.}, Cu$^{1+}$ or Ag$^{1+}$), the magnetic sublattices formed in $M^{1+}M'^{3+}$P$_2$S$_6$ extend the magnetic structures of the usually magnetically hexagonal $M_2$P$_2$S$_6$ compounds by an alternating/triangular and a zig-zag stripe-like magnetic arrangement~\cite{PColombet1982,PColombet1983,RBrec1986}. The stripe-like magnetic structure is especially notable, as each stripe of magnetic ions is well isolated from the adjacent magnetic stripes by a stripe of non-magnetic ions. Although the corresponding compound still has a (quasi-)2D layered crystal structure, the magnetic structure can be expected to exhibit 1D magnetic characteristics. Indeed, several indications for such low dimensional magnetism are reported for $M^{1+}M'^{3+}$P$_2$S$_6$ with $M = \textrm{Ag}$ and $M' = \textrm{Cr}$~\cite{PColombet1983,HMutka1993}, making it an interesting compound for further studies.

However, until now only details on the synthesis of AgCrP$_2$S$_6$ \textit{via} solid state synthesis are reported\footnote{Although Mutka \textit{et al.}~\cite{HMutka1993} mention CVT grown crystals, they do not report any further details or conditions regarding the crystal growth.} \cite{PColombet1983}. Although small crystals in the $\mu$m scale could be obtained by solid state synthesis, which allowed for a structural solution based on single crystal X-ray diffraction, significantly larger crystals are needed for detailed investigations of the physical properties including anisotropies. Thus we have optimized  the crystal growth conditions of AgCrP$_2$S$_6$ by the chemical vapor transport (CVT) technique, in a similar approach to the successful single crystal growth of most ternary $M_2$P$_2$S$_6$ compounds using this technique~\cite{BTaylor1973,Selter2021}. Hereafter, the optimized conditions for the single crystal growth of the quarternary compound AgCrP$_2$S$_6$ are reported. Furthermore, a comprehensive compositional and structural characterization of the as-grown crystals is presented.

\section{Materials \& Methods}

The elemental educts for the crystal growth of AgCrP$_2$S$_6$, as listed in Tab.~\ref{tab:educts}, were obtained from Alfa Aesar and kept in an argon filled glove box for storage and handling.

\begin{table}[htb]
    \centering
    \begin{tabular}{lcc}
    \toprule
    Chemical & Specification & Purity \\
    \midrule
    Silver & Powder, APS 4--7 micron & 99.9\%  \\
    Chromium & Powder, -100+325 mesh & 99.99\% \\
    Red phosphorus & Lumps & 99.999\% \\
    Sulfur & Pieces & 99.999\% \\
    Iodine & Resublimed crystals & 99.9985\% \\
    \bottomrule
    \end{tabular}
    \caption{Elemental educts used for the CVT growth of AgCrP$_2$S$_6$.}
    \label{tab:educts}
\end{table}

The crystals obtained from the CVT crystal growth experiments were thoroughly characterized by scanning electron microscopy (SEM) regarding their morphology and topography using a secondary electron (SE) detector and regarding chemical homogeneity \textit{via} the chemical contrast obtained from a back scattered electron (BSE) detector. For this, a ZEISS EVO MA 10 scanning electron microscope was used. The chemical composition of the crystals was investigated by energy dispersive X-ray spectroscopy (EDX), which was measured in the same SEM device with an accelerating voltage of 30\,kV for the electron beam and using an energy dispersive X-ray analyzer.

The crystal structure of the obtained crystals was investigated by powder X-ray diffraction (pXRD), which was measured on a STOE STADI laboratory diffractometer in transmission geometry with Cu-K$_{\alpha1}$ radiation from a curved Ge(111) single crystal monochromator and detected by a MYTHEN 1K 12.5$^\circ$-linear position sensitive detector manufactured by DECTRIS. The pXRD patterns were initially analyzed by pattern matching using the HighScore Plus program suite~\cite{Degen2014}. After the crystallographic phase was identified, a structural refinement of the crystal structure model was performed based on our experimental patterns using the Rietveld method in Jana2006~\cite{Petricek2014}.

\section{Crystal Growth \textit{via} Chemical Vapor Transport}

All procedures for the preparation were performed under argon atmosphere in a glove box, the elemental educts silver, chromium , red phosphorus and sulfur were weighed out in a molar ratio of Ag\,:\,Cr\,:\,P\,:\,S~=~1\,:\,1\,:\,2\,:\,6 and homogenized in an agate mortar. 0.5\,g of reaction mixture were loaded in a quartz ampule (6\,mm inner diameter, 2\,mm wall thickness) together with approx. 50\,mg of the transport agent iodine. Immediately prior to use, the ampule was cleaned by washing with distilled water, rinsing with isopropanol and, subsequently, baking out at 800\,\gc\ for at least 12\,h in an electric tube furnace. This is done to avoid contamination of the reaction volume with (adsorbed) water. The filled ampule was then transferred to a vacuum pump and evacuated to a residual pressure of $10^{-8}$\,bar. To suppress the unintended sublimation of the transport agent during evacuation, the end of the ampule containing the material was cooled with a small Dewar flask filled with liquid nitrogen. After reaching the desired internal pressure, the valve to the vacuum pump was closed, the cooling was stopped and the ampule was sealed under static pressure at a length of approximately 12\,cm.

\begin{figure}[htb]
\centering
\includegraphics[width=\columnwidth]{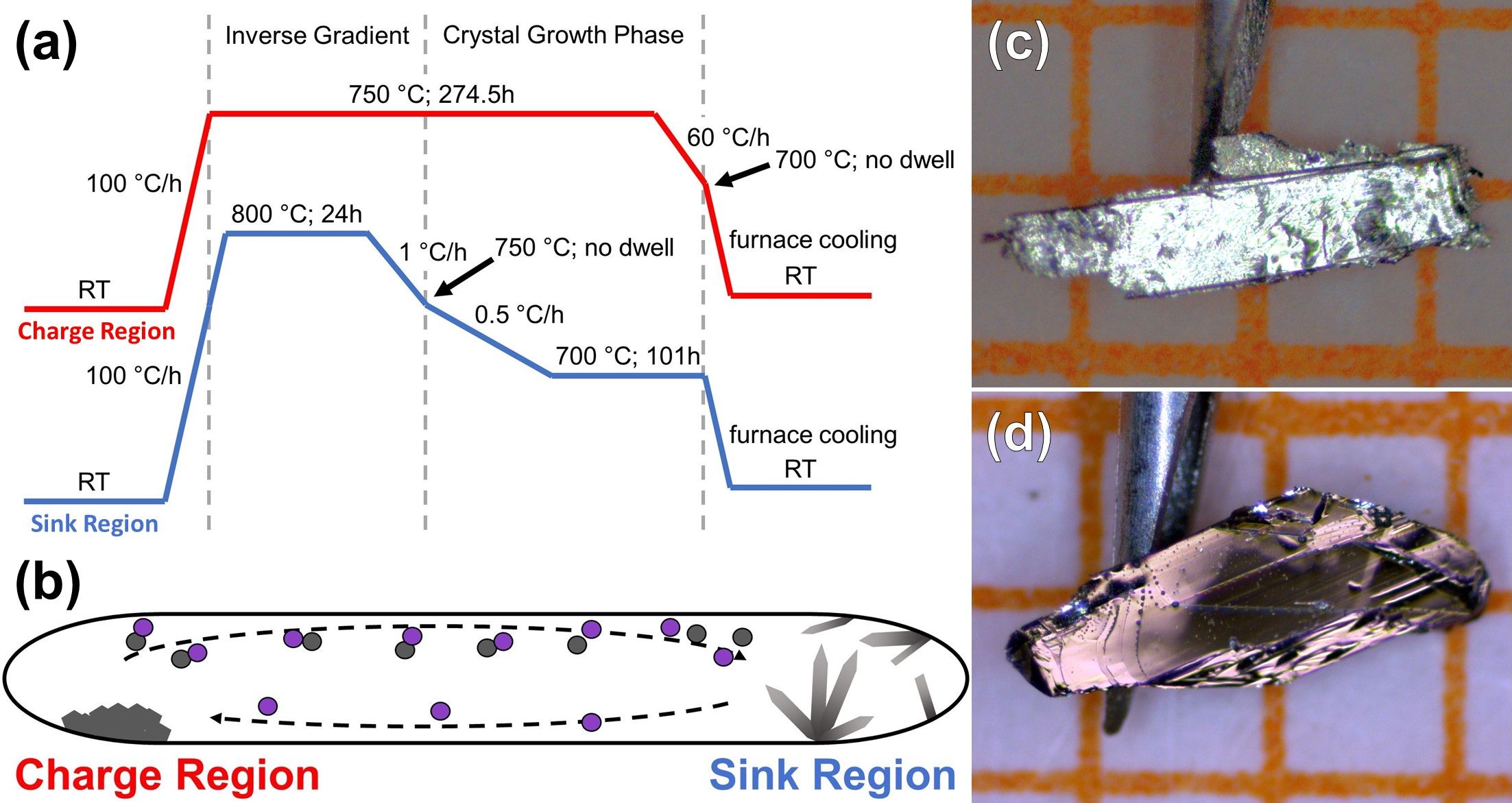}
\caption{
(a) Graphical illustrations of the temperature profile for the CVT growth of AgCrP$_2$S$_6$ and (b) schematic drawing of an ampule during CVT. Arrows indicate the mass flow of the volatile transport species (top) and the flow of the released transport agent back to the charge (bottom). (c) and (d): As-grown crystals of AgCrP$_2$S$_6$. A orange square in the background corresponds to $1\,\textrm{mm} \times 1\,\textrm{mm}$ for scale.
}
\label{fig:Tprofile_CrystalImage}
\end{figure}

%\begin{figure}[htb]
%\centering
%\includegraphics[width=.7\columnwidth]{T_profile.JPG}
%\caption{
%Graphical illustration of the temperature profile for the CVT growth of AgCrP$_2$S$_6$.
%}
%\label{fig:Tprofile}
%\end{figure}

The ampule was carefully placed in a two-zone tube furnace in such a way that the reaction mixture was only at one side of the ampule which is referred to as the charge region. As illustrated in Fig.~\ref{fig:Tprofile_CrystalImage}(a), the furnace was initially heated homogeneously to 750\,\gc\ at 100\,\gc/h. The charge region was kept at this temperature for 274.5\,h while the other side of the ampule, which is the sink region (see Fig.~\ref{fig:Tprofile_CrystalImage}(b)), was initially heated up to 800\,\gc\ at 100\,\gc/h, dwelled at this temperature for 24\,h and then cooled back to 750\,\gc\ at 1\,\gc/h. An inverse transport gradient is formed, \textit{i.e.} transport from sink to charge, to clean the sink region of particles which stuck to the walls of the quartz ampule during the previous preparation steps. This ensures improved nucleation conditions in the following step. Then the sink region was cooled to 690\,\gc\ at 0.5\,\gc/h to gradually form the thermal transport gradient resulting in a controlled nucleation. With a final gradient of 750\,\gc\ (charge) to 690\,\gc\ (sink), the ampule was dwelled for 100\,h. After this period of time, the charge region was cooled to the sink temperature in 1\,h before both regions were furnace cooled (\textit{i.e.}, the heating elements were turned off) to room temperature.

%\begin{figure}[htb]
%\centering
%\includegraphics[width=.7\columnwidth]{Crystal_Image.JPG}
%\caption{
%Photographic image of an as-grown crystal of AgCrP$_2$S$_6$. A orange square in the background corresponds to $1\,\textrm{mm} \times %1\,\textrm{mm}$ for scale.
%}
%\label{fig:CrystalImage}
%\end{figure}

Shiny plate-like crystals of AgCrP$_2$S$_6$ in the size of approximately $2\,\textrm{mm} \times 2\,\textrm{mm} \times 100\,\mu\textrm{m}$ were obtained. As example, as-grown single crystals are shown in Fig.~\ref{fig:Tprofile_CrystalImage}(c) and (d). These crystals exhibit a layered morphology and are easily exfoliated, which is typical for bulk crystals of (quasi-)2D materials. On the surface of some crystals, small spheres (likely solidified droplets) of a secondary phase were found. As this secondary phase has been formed purely on the surface, exfoliation of the crystal was suitable to remove this impurity phase.

\section{Crystal Morphology \& Compositional Analysis}

\begin{figure}[htb]
\centering
\includegraphics[width=\columnwidth]{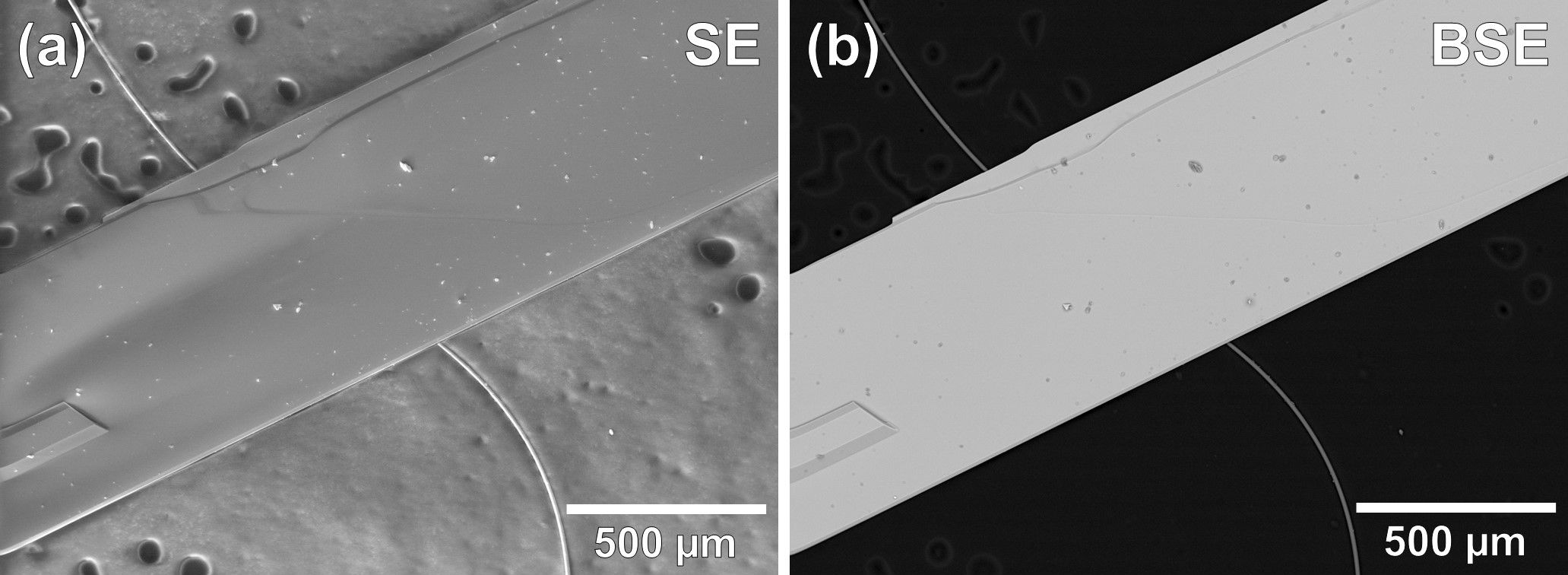}
\caption{
SEM image of an as-grown AgCrP$_2$S$_6$ crystal with topographical contrast (SE mode) in (a) and chemical contrast (BSE mode) in (b).
}
\label{fig:SEM_SE_BSE}
\end{figure}

The topographical SE image of an as grown AgCrP$_2$S$_6$ crystal in Fig.~\ref{fig:SEM_SE_BSE}(a) exhibits a flat crystal surface and sharp edges. The terrace close to the upper edge of the crystal is a typical feature of layered systems. Also on the upper edge of the crystal, some steps can be seen, which form 120$^\circ$ angles, indicative of a hexagonal crystal habitus. The SE image with BSE detector in in Fig.~\ref{fig:SEM_SE_BSE}(b) shows an overall homogeneous contrast over the surface of the crystal demonstrating that it is chemically homogeneous. At some small areas, a change in contrast is observed. In comparison with the SEM image in SE mode, these spots can be clearly attributed to impurity particles on top of the crystal and not to any region of intergrowth with a secondary phase.

By EDX measurements on multiple spots on several crystals, the mean elemental composition of the crystals was obtained as Ag$_{1.03(2)}$Cr$_{1.06(2)}$P$_{2.03(2)}$S$_{5.88(1)}$. This composition is in ideal agreement with the expected composition of AgCrP$_{2}$S$_{6}$ and the small standard deviations (given in parentheses) indicate a homogeneous elemental distribution and composition.

\section{Structural Analysis}

%pXRD was measured on pulverized crystals of AgCrP$_2$S$_6$ which were exfoliated before grinding to avoid secondary phase contributions to the pattern. As shown in Fig.~\ref{fig:AgCrPS_pXRD}(a), using this procedure a phase pure pattern of AgCrP$_2$S$_6$ was obtained.

\begin{figure}[htb]
\centering
\includegraphics[width=\columnwidth]{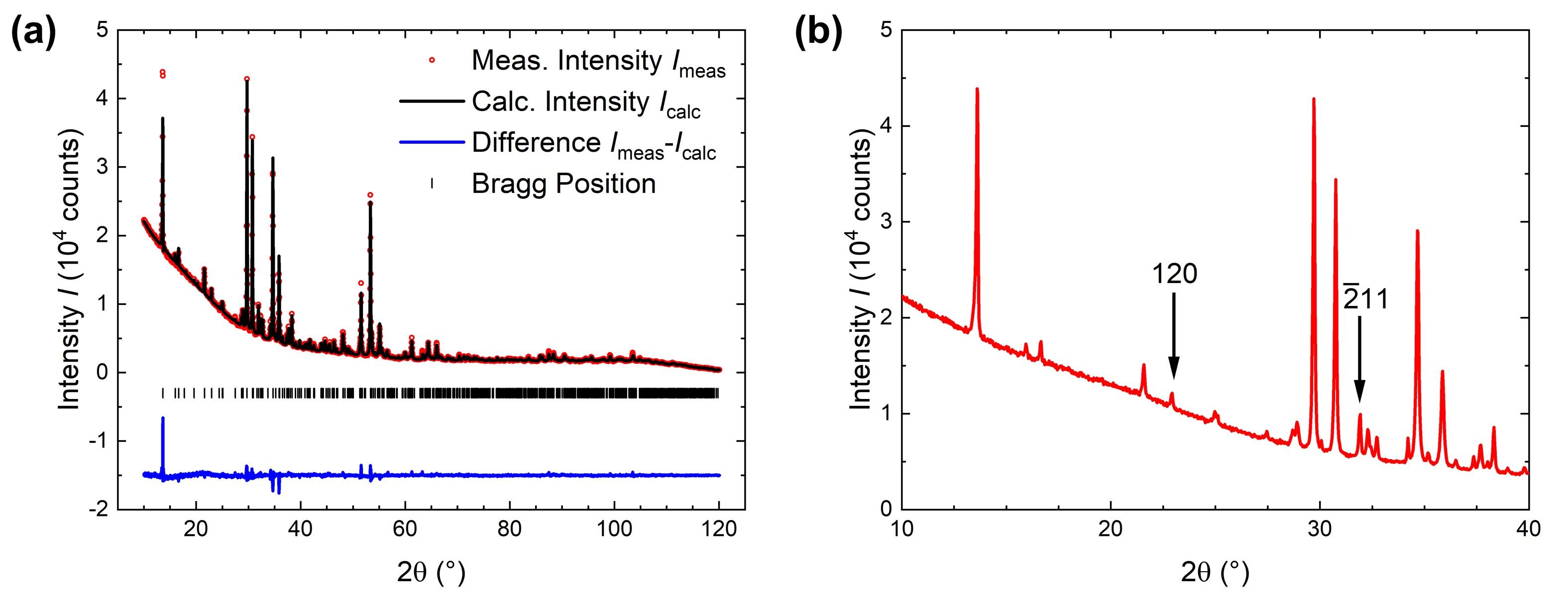}
\caption{
(a) pXRD pattern from powdered AgCrP$_2$S$_6$ crystals compared to the calculated pattern based on the refined crystal structure model. (b) Zoomed-in view on the low angle $2\theta$ regime (10--40$^\circ$). The marked reflections are expected to be systematically absent assuming a space group of $C2/m$ instead of $P2/a$.
}
\label{fig:pXRD}
\end{figure}

The pXRD pattern obtained from AgCrP$_2$S$_6$ crystals, as shown in Fig.~\ref{fig:pXRD}(a), was indexed in the space group $P2/a$ (No.~13), in agreement with literature~\cite{PColombet1983}. No additional reflection were observed demonstrating the phase purity of our crystals.

The $C2/m$ space group, which is typically observed for compounds of the $M_2$P$_2$S$_6$ family~\cite{BTaylor1973,RBrec1980}, including $M^{2+}M'^{2+}$P$_2$S$_6$ compounds of isovalent substitution series (\textit{e.g.}, MnFeP$_2$S$_6$~\cite{TMasubuchi2008}, MnNiP$_2$S$_6$~\cite{Shemerliuk2021} and FeNiP$_2$S$_6$~\cite{Selter2021}), can be ruled out. Assuming a monoclinic unit cell, several observed reflections correspond to Laue indices that are systematically absent for $C$ centering, as they violate the reflection condition $hkl$: $h + k = 2n$. Examples are the reflections at $2\theta = 22.94^\circ$ corresponding to $120$ and at  $2\theta = 31.92^\circ$ corresponding to $\overline{2}11$, as shown in Fig.~\ref{fig:pXRD}(b).

This implies that Ag and Cr indeed arrange as zig-zag stripes in AgCrP$_2$S$_6$ and do not just randomly occupy the corners of the structural honeycomb network, as it is the case for isovalent substitutions. While the former scenario breaks the mirror symmetry of the $C2/m$ space group of the Fe$_2$P$_2$S$_6$ aristotype~\cite{WKlingen1970}, which results in a $P2/a$ space group, the latter scenario would not. Also a $C2/c$ space group, as reported, \textit{e.g.}, for CuCrP$_2$S$_6$~\cite{PColombet1982} with a triangular arrangement of the two transition element cations, can be ruled out based on the same considerations. 

%\begin{table}[tb]
 %\caption{
 %Summary of experimental parameters of the pXRD experiment on AgCrP$_2$S$_6$, extracted lattice parameters and reliability factors of the structural model obtained by the Rietveld method.
 %}
 %\centering
 %\begin{tabular}{lc}
 %\hline \hline
 %\multicolumn{2}{l}{\textit{Experiment}} \\
 %Temperature (K) & 293(2) \\
 %Radiation & Cu-K$_{\alpha_1}$ \\
 %Wavelength (\r{A}) & 1.54059 \\
 %$\theta_\textrm{min}$ ($^\circ$) & 10.00 \\
 %$\theta_\textrm{step}$ ($^\circ$) & 0.03 \\ 
 %$\theta_\textrm{max}$ ($^\circ$) & 120.13 \\
 %&\\
 %\multicolumn{2}{l}{\textit{Crystal Data}} \\
 %Crystal System & Monoclinic \\
 %Space Group & $P2/a$ \\
 %$a$ (\r{A}) & 5.8832(1) \\
 %$b$ (\r{A}) & 10.6214(2) \\
 %$c$ (\r{A}) & 6.7450(3) \\
 %$\beta$ ($^\circ$) & 106.043(2) \\
 %&\\
 %\multicolumn{2}{l}{\textit{Refinement}} \\
 %Goodness-Of-Fit & 2.13 \\
 %$R_\textrm{p}$ (\%) & 2.08 \\
 %$wR_\textrm{p}$ (\%) & 3.11 \\
 %$R_\textrm{F}$ (\%) & 5.45 \\
 %\hline \hline
 %\end{tabular}
 %\label{tab:summary_pXRD}
%\end{table}

\begin{table}[tb]
 \caption{
  Top: Summary of experimental parameters of the pXRD experiment on AgCrP$_2$S$_6$, extracted lattice parameters and reliability factors of the structural model obtained by the Rietveld method. Bottom: Refined crystal structure model of AgCrP$_2$S$_6$ and isotropic displacement parameters with standard deviations given in parentheses. All sites were treated as fully occupied.
 }
 \centering
 %\resizebox{\columnwidth{}}{!}{
 \begin{tabular}{ccccccc}
 \toprule
 \multicolumn{4}{c}{\textit{Experiment}} &&& \\
 \multicolumn{4}{c}{Temperature (K)} & \multicolumn{3}{c}{293(2)} \\
 \multicolumn{4}{c}{Radiation} & \multicolumn{3}{c}{Cu-K$_{\alpha_1}$} \\
 \multicolumn{4}{c}{Wavelength (\r{A})} & \multicolumn{3}{c}{1.54059} \\
 \multicolumn{4}{c}{$\theta_\textrm{min}$ ($^\circ$)} & \multicolumn{3}{c}{10.00} \\
 \multicolumn{4}{c}{$\theta_\textrm{step}$ ($^\circ$)} & \multicolumn{3}{c}{0.03} \\ 
 \multicolumn{4}{c}{$\theta_\textrm{max}$ ($^\circ$)} & \multicolumn{3}{c}{120.13} \\
 &&&&&\\
 \multicolumn{4}{c}{\textit{Crystal Data}} &&& \\
 \multicolumn{4}{c}{Crystal System} & \multicolumn{3}{c}{monoclinic} \\
 \multicolumn{4}{c}{Space Group} & \multicolumn{3}{c}{$P2/a$} \\
 \multicolumn{4}{c}{$a$ (\r{A})} & \multicolumn{3}{c}{5.8832(1)} \\
 \multicolumn{4}{c}{$b$ (\r{A})} & \multicolumn{3}{c}{10.6214(2)} \\
 \multicolumn{4}{c}{$c$ (\r{A})} & \multicolumn{3}{c}{6.7450(3)} \\
 \multicolumn{4}{c}{$\beta$ ($^\circ$)} & \multicolumn{3}{c}{106.043(2)} \\
 &&&&&\\
 \multicolumn{4}{c}{\textit{Refinement}} &&& \\
 \multicolumn{4}{c}{Goodness-Of-Fit} & \multicolumn{3}{c}{2.13} \\
 \multicolumn{4}{c}{$R_\textrm{p}$ (\%)} & \multicolumn{3}{c}{2.08} \\
 \multicolumn{4}{c}{$wR_\textrm{p}$ (\%)} & \multicolumn{3}{c}{3.11} \\
 \multicolumn{4}{c}{$R_\textrm{F}$ (\%)} & \multicolumn{3}{c}{5.45} \\
 \bottomrule
 &&&&&\\
 \toprule
 \multirow{2}{*}{Label} & \multirow{2}{*}{Type} & \multirow{2}{*}{Wyck} & \multirow{2}{*}{$x$} & \multirow{2}{*}{$y$} & \multirow{2}{*}{$z$} & $U_\textrm{iso}$ \\
  & & & & & & ($\times 10^{-3} \textrm{\r{A}}^2$) \\
 \midrule
 Ag1 & Ag & $2e$  & 0.75      & 0.4364(2) & 0         & 34(1) \\
 Cr1 & Cr & $2e$  & 0.25      & 0.9229(4) & 0         & 20(2) \\
 P1  & P  & $4g$  & 0.2979(6) & 0.2466(4) & 0.1659(6) &  2(1) \\
 S1  & S  & $4g$  & 0.9792(7) & 0.2309(5) & 0.2336(8) & 18(2) \\
 S2  & S  & $4g$  & 0.9880(5) & 0.9233(4) & 0.2165(8) &  7(1) \\
 S3  & S  & $4g$  & 0.4777(7) & 0.3947(4) & 0.2802(7) & 34(1) \\
 \bottomrule
 \end{tabular}
 %}
 \label{tab:summary_pXRD}
\end{table}

\begin{figure}[htb]
\centering
\includegraphics[width=\columnwidth]{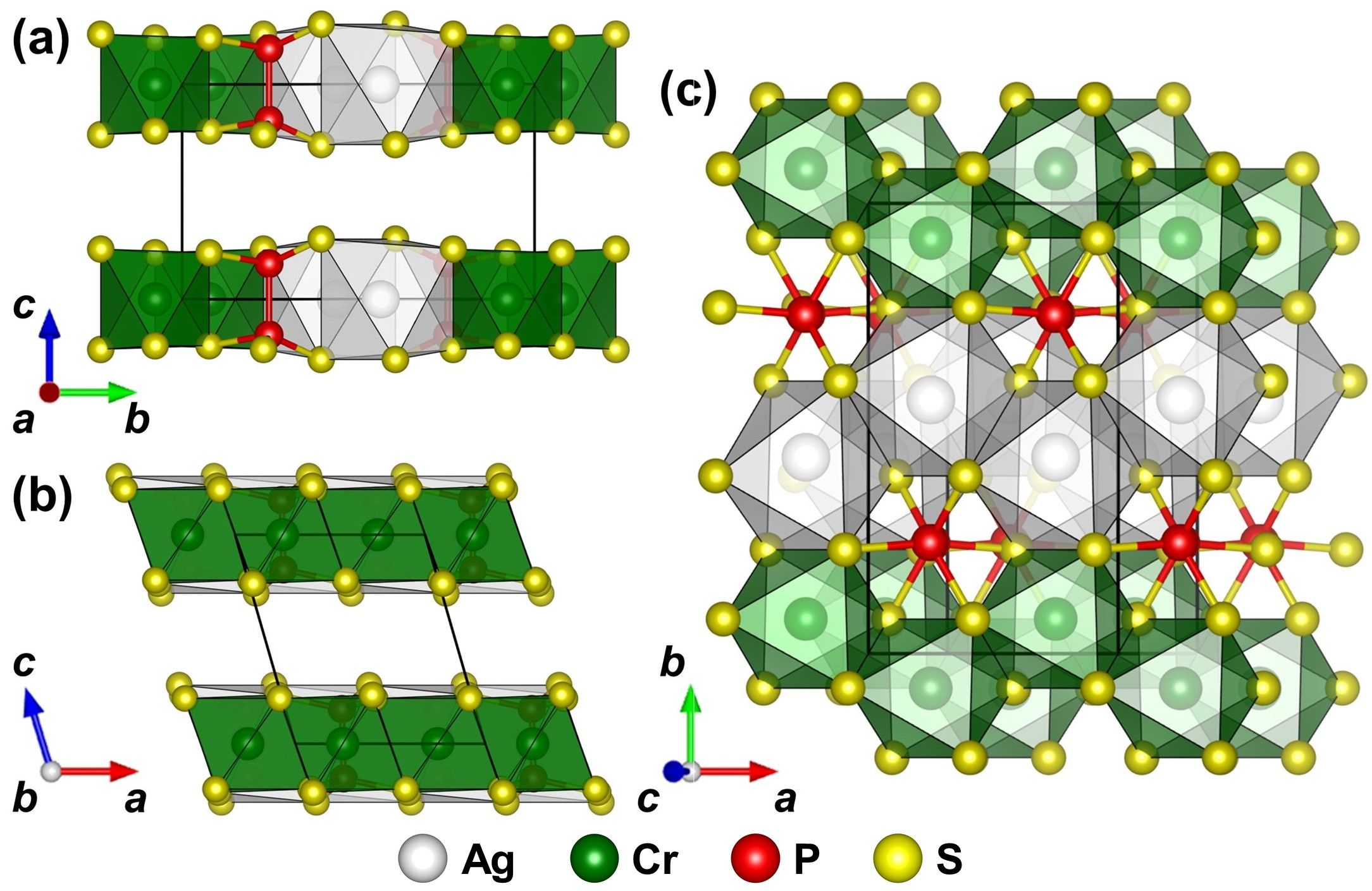}
\caption{
Refined crystal structure model of AgCrP$_2$S$_6$ after Rietveld refinement. View along $a$ in (a), along $b$ in (b) and along $c^*$ in (c). The CrS$_6$ and AgS$_6$ coordination environments are shown in the color of the respective central atom.
}
\label{fig:structure}
\end{figure}

Starting from the crystal structure model proposed by Colombet \textit{et al.}~\cite{PColombet1983}, a refined crystal structure model is obtained using the Rietveld method which is sufficient to describe our experimental pattern with good agreement, as shown in Fig.~\ref{fig:pXRD}(a). The obtained lattice parameter and reliability factors are summarized in Tab.~\ref{tab:summary_pXRD} (top) and the refined structural model is given in the same table on the bottom and is illustrated in Fig.~\ref{fig:structure}. The strongest disagreement between model and experiment is observed for the high intensity $001$ reflection at $2\theta = 13.67^\circ$. As this reflection corresponds to the stacking of layers, it is most prominently affected by any kind of disorder or defects influencing the stacking. Due to the weak structural interaction between layers, which are only based on weak van der Waals forces, the $M_2$P$_2$S$_6$ compounds are prone to stacking faults and twinning between layers. In the presence of such defects, the shape of the corresponding $001$ reflection is altered, which may be a reason for the observed deviation between the experiment and the model without defects.

Additionally, the experimental pattern exhibits significantly altered reflection intensities compared to an initial model, which are attributed to a strongly preferred orientation of the crystallites in the investigated sample. Due to the layered structure with only weak van der Waals interactions between layers, the powder particles obtained from grinding AgCrP$_2$S$_6$ crystals are plate-like and tend to lie flat on the sample holder. Thus, reflections with a dominant $l$ component (\textit{e.g.}, $001$) exhibit higher intensities than expected for spherical crystallites in transmission geometry. To adjust for this effect in the model, the method proposed by March~\cite{AMarch1932} and extended by Dollase~\cite{WDollase1986} was used. However, the preferred orientation in AgCrP$_2$S$_6$ is strongly pronounced, such that it might be beyond the limit of what the semi-empirical March-Dollase model is capable of describing accurately. This may furthermore contribute to the deviation between model and experiment around the $001$ reflection.

The refined crystal structure model for AgCrP$_2$S$_6$ shows that the Ag--S bonds are notably longer than the Cr--S bonds, as expected based on the difference between the size of the transition element cations (\textit{e.g.}, ionic radii for octahedral coordination: $r(\textrm{Ag}^{1+}) = 1.15$\,\r{A} and $r(\textrm{Cr}^{3+}) = 0.62$\,\r{A}~\cite{RShannon1976}). These different bond lengths result in a distortion of the structure compared to the aristotype Fe$_2$P$_2$S$_6$, which can be clearly observed, \textit{e.g.}, in Fig.~\ref{fig:structure}(a). In detail, the CrS$_6$ coordination environment remains antiprismatic (i.e., close to octahedral with a slight trigonal elongation along $c^*$) with the faces above and below the shared plane of the transition elements being parallel to each other. However, the AgS$_6$ coordination environment as well as for the P$_2$S$_6$ environment are distorted in such a way that the faces above and below the transition element plane are not parallel to each other. In the view along the $c^*$ direction in Fig.~\ref{fig:structure}(c), this distortion manifests in Ag and P$_2$ being shifted off-center in their respective sulfur coordination environments away from the closest Cr positions. Meanwhile, Cr is located exactly in the center of the CrS$_6$ unit. The observation that the CrS$_6$ unit is closer to an ideal octahedral coordination environment than the AgS$_6$ unit can be understood considering the local charge density (\textit{i.e.}, ionic size and charge). Cr$^{3+}$ is small and highly charged and, thus, interacts with the surrounding S atoms stronger than the comparable large and less charged Ag$^{1+}$. Another notable structural aspect is the strong distortion of the [P$_2$S$_6$]$^{4-}$ units that demonstrates how flexible this covalent complex anion is. This complex anion is a common and characteristic building unit in the $M_2$P$_2$S$_6$ family and its flexibility may indicate that several more compounds of the general formula $M^{1+}M'^{3+}$P$_2$S$_6$ are stable but have not been synthesized yet.

\section{Summary \& Conclusions}

We report optimized crystal growth conditions for the quarternary compound AgCrP$_2$S$_6$ \textit{via} Chemical Vapor Transport (CVT). A temperature profile adapted from the CVT growth of ternary $M_2$P$_2$S$_6$ compounds is sufficient to yield crystals of the target AgCrP$_2$S$_6$ phase in the mm-size. On some crystals, traces of a superficial impurity phase is found which could be readily removed by exfoliation.

The as-grown crystals exhibit a plate-like, layered morphology as well as a hexagonal habitus and have the expected composition of AgCrP$_2$S$_6$ based on EDX spectroscopy. The pXRD pattern is indexed in the space group $P2/a$ in agreement with literature~\cite{PColombet1983}. The $P2/a$ space group, on which the zig-zag type arrangement of $M$ and $M'$ is based on, can be well distinguished from, \textit{e.g.}, the $C2/m$ and $C2/c$ space groups due to reflections that are systematically absent for $C$ centering. Starting from the model of Colombet \textit{et al.}~\cite{PColombet1983}, a refined structural model is obtained using the Rietveld method. This model contains a notable distortion of the AgS$_6$ and P$_2$S$_6$ coordination environments, while the CrS$_6$ units remain antiprismatic with a slight trigonal distortion.

The zig-zag stripe-like arrangement in AgCrP$_2$S$_6$ and the alternating arrangement of $M$ and $M'$, which is reported, \textit{e.g.}, for CuCrP$_2$S$_6$, are promising to yield interesting magnetic and electronic structures. While only few such quarternary phosphorus sulfide compounds have been synthesized until now, many more combinations of a 1+-ion and a 3+-ion can be expected to form analogous compounds. Furthermore, the fundamental idea of replacing $M^{X+}$ by $M^{(X-1)+}_{0.5}M'^{(X+1)+}_{0.5}$ may be adoptable to the closely related structures such as, $M^{3+}_2$(Si,Ge)$_2$Te$_6$ compounds.

The single crystals of AgCrP$_2$S$_6$ that were obtained using the presented growth conditions allow for studies of the low dimensional magnetic interactions including the magnetic anisotropy of this compound in the future, which may lead to a better fundamental understanding of low dimensional magnetism. Furthermore, the van der Waals layered structure makes exfoliation easily possible and, thus, our successful growth of single crystals paves the way for further manufacturing of few-layer or even monolayer samples of AgCrP$_2$S$_6$.

\vspace{6pt} 

\funding{This work is supported by the Deutsche Forschungsgemeinschaft (DFG) via Grant No.~DFG~A.S~523\textbackslash4-1. S.S. acknowledges financial support from GRK-1621 graduate academy of the DFG. B.B. acknowledges financial support from the DFG through SFB~1143~(project-id 247310070). Y.S acknowledge the support of BMBF through UKRATOP (BMBF). S.A., B.B. and S.S. thank DFG for financial support in the frame of the joint DFG-RSF project-id 405940956.}

\dataavailability{The refined crystal structure model and the powder X-ray diffraction dataset of AgCrP$_2$S$_6$ presented in this study are openly available in the Crystallography Open Database (COD), COD ID: 3000295 under \url{https://www.crystallography.net/cod/3000295.html}.}

%\acknowledgments{}

\conflictsofinterest{The authors declare no conflict of interest.}

\sampleavailability{Single crystals of AgCrP$_2$S$_6$ are available from the corresponding author.}

\abbreviations{Abbreviations}{The following abbreviations are used in this manuscript:\\

\noindent 
\begin{tabular}{@{}ll}
2D & Two-dimensional\\
CVT & Chemical vapor transport\\
SEM & Scanning electron microscopy\\
SE & Secondary electron\\
BSE & Back-scattered electron\\
EDX & Energy dispersive X-ray spectroscopy\\
pXRD & Powder X-ray diffraction\\

\end{tabular}}

\end{paracol}
\reftitle{References}

\externalbibliography{yes}
\bibliography{literature}

\end{document}